\documentclass[a4paper,11pt]{article}
\usepackage{pos}
\usepackage{caption}
\usepackage{subcaption}

\title{Comparison of the measured atmospheric muon rate with Monte Carlo simulations and sensitivity study for detection of prompt atmospheric muons with KM3NeT}
 \ShortTitle{Measured and simulated atmospheric muons and sensitivity to prompt muons}

\author*[a]{Piotr Kalaczyński}

\affiliation[a]{National Centre for Nuclear Research,\\
  Pasteura 7, Warsaw, Poland}

\forColl{KM3NeT} 

\emailAdd{piotr.kalaczynski@ncbj.gov.pl}

\abstract{The KM3NeT Collaboration has successfully deployed the first detection units of the next generation undersea neutrino telescopes in the Mediterranean Sea at the two sites in Italy and in France. The data sample collected between December 2016 and January 2020 has been used to measure the atmospheric muon rate at two different depths under the sea level: 3.5 km with KM3NeT-ARCA and 2.5 km with KM3NeT-ORCA. Atmospheric muons represent an abundant signal in a neutrino telescope and can be used to test the reliability of the Monte Carlo simulation chain and to study the physics of extensive air showers caused by highly-energetic primary nuclei impinging the Earth’s atmosphere. At energies above PeV the contribution from prompt muons, created right after the first interaction in the shower, is expected to become dominant, however its existence has not yet been experimentally confirmed. In this talk, data collected with the first detection units of KM3NeT are compared to Monte Carlo simulations based on MUPAGE and CORSIKA codes. The main features of the simulation and reconstruction chains are presented. Additionally, the first results of the simulated signal from the prompt muon component for KM3NeT-ARCA and KM3NeT-ORCA obtained with CORSIKA  are discussed.}

\FullConference{37$^{\rm{th}}$ International Cosmic Ray Conference (ICRC 2021)\\
		July 12th -- 23rd, 2021\\
		Online -- Berlin, Germany}

\begin{document}
\maketitle

\section{Introduction}

KM3NeT is a research infrastructure under construction at the bottom of the Mediterranean Sea. It consists of two neutrino detectors: ARCA (Astroparticle Research with Cosmics in the Abyss) located off-shore Portopalo di Capo Passero, Sicily, Italy, at a depth of 3500\,m and ORCA (Oscillation Research with Cosmics in the Abyss) off-shore Toulon, France, at a depth of 2450\,m.

The main goal of ARCA is to detect TeV-PeV neutrinos from astrophysical sources or in coincidence with other high energy events, for example gravitational waves, gamma ray bursts or blazar flares. The ORCA detector will study oscillations of the atmospheric neutrinos in the GeV range, in order to determine the neutrino mass hierarchy (NMH) \cite{LoI}.

Both detectors have the same structure. They consist of vertically aligned detection units (DUs), each carrying 18 digital optical modules (DOMs) \cite{LoI}. Each DOM houses 31 3-inch photomultiplier tubes (PMTs), calibration and positioning instruments and readout electronics boards. ARCA and ORCA differ in the horizontal (90\,m and 20\,m respectively) and vertical (36\,m and 9\,m respectively) spacing between the DOMs, which is optimised for the energy ranges mentioned above. Currently, both detectors are taking data with 6 DUs installed at each site. In final configurations, there will be 115 DUs at ORCA and 2x115 DUs (in two blocks) at ARCA site. 

Atmospheric muons are the most abundant signal in a neutrino telescope and a major background to the physics analyses with neutrinos, however they also prove useful in testing the detector performance and validity of the simulations. In this paper, a comparison of the first data collected by ARCA and ORCA to the Monte Carlo (MC) simulations is presented.

Muon data contains valuable information about the extensive air showers (EAS) forming in the Earth's atmosphere and about the cosmic ray (CR) primaries that cause them. At muon energies above PeV, a contribution from muons created by quickly decaying heavy hadrons (often containing the charm quark), called prompt muons, is predicted. It has not been experimentally confirmed yet. The first results evaluating the expected signal due to the prompt muon contribution are shown in secion \ref{prompt-ana}. They are based on muon simulations for the full ARCA and ORCA detectors.

\section{Muon simulation}\label{mu_sim}

The KM3NeT simulation chain is divided into several stages. The generation of the atmospheric muon bundles is performed using two different codes: MUPAGE \cite{MUPAGE} and CORSIKA \cite{CORSIKA}. MUPAGE is a fast simulation program, based on parametric formulas that generates muon bundles, induced by CRs impinging the Earth’s atmosphere. The muon distributions are sampled on the surface of the active volume of the detector at different undersea depths and zenith angles. CORSIKA is a software package that simulates the interactions of the primary CRs in the upper atmosphere and follows the development of the shower to a specified observation level (here, it is the sea level). The CORSIKA package is flexible and allows to choose between different models to describe the hadronic interactions and the composition of the primary CR flux.

The GENIE-based code, gSeaGen  \cite{gSeaGen,GENIE} is a multipurpose tool, commonly used in KM3NeT to simulate the neutrino interactions in the detectors. In this work, gSeaGen is used to convert the CORSIKA output into the KM3NeT standard format, compute event weights according to the specified CR composition model and propagate muon bundles from the sea level to the active volume of the detector. The propagation is done using a 3-dimensional muon propagator, in this case: PROPOSAL \cite{PROPOSAL}. This step is not necessary for MUPAGE, as it already generates events at the detector level. 

After the propagation, Cherenkov light emission along the path of muons and their probability to be detected by the DOMs are simulated with a custom application developed for KM3NeT called JSirene, using multidimensional interpolation tables \cite{JSirene}. Subsequently, the following effects expected in the real data are also included in the simulated set: the environmental optical background due to the bioluminescence and the \(^{40}\mathsf{K}\) decay,  detector response,
taking into account the front-end electronics. For MUPAGE the simulation is performed in a run-by-run mode, which means that each data taking run is simulated separately. On the other hand, for CORSIKA a muon flux averaged over all runs is simulated. The same trigger algorithms as used for real data are applied to identify possible interesting events in the simulated sample \cite{AtmMuRate}. Finally, the selected events are processed with the track reconstruction program "JGandalf", used for the real data stream as well \cite{reco}. At this stage the simulated MC events can be easily compared to the calibrated data. A  track reconstruction algorithm is applied both to data and MC events.

\section{Data vs MC comparisons}\label{data-MC}

\begin{figure}
     \centering
     \begin{subfigure}[b]{0.49\textwidth}
         \centering
         \includegraphics[width=\textwidth]{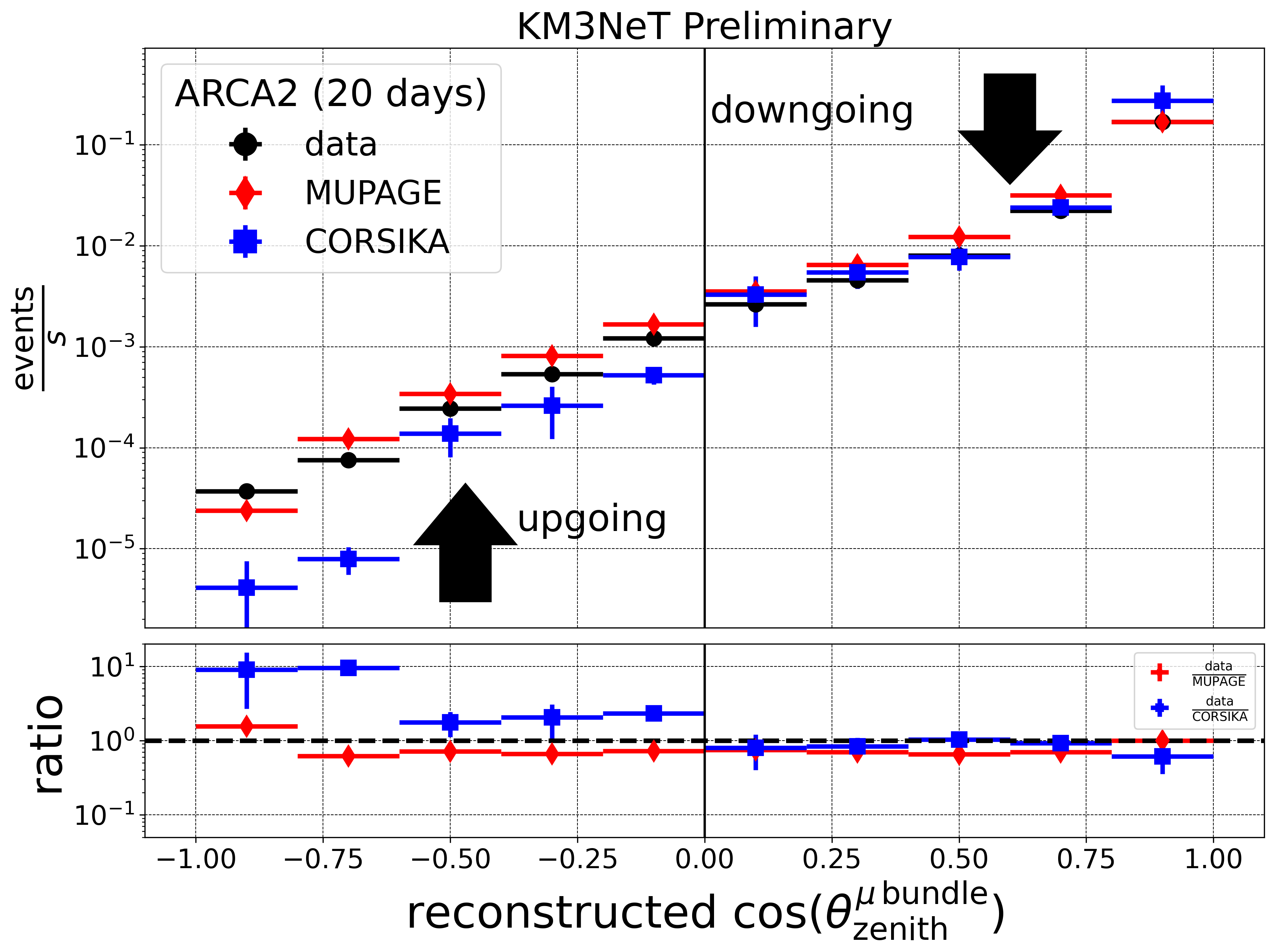}
         \caption{Distribution for ARCA2.}
         \label{fig:zenith-1}
     \end{subfigure}
     \hfill
     \begin{subfigure}[b]{0.49\textwidth}
         \centering
         \includegraphics[width=\textwidth]{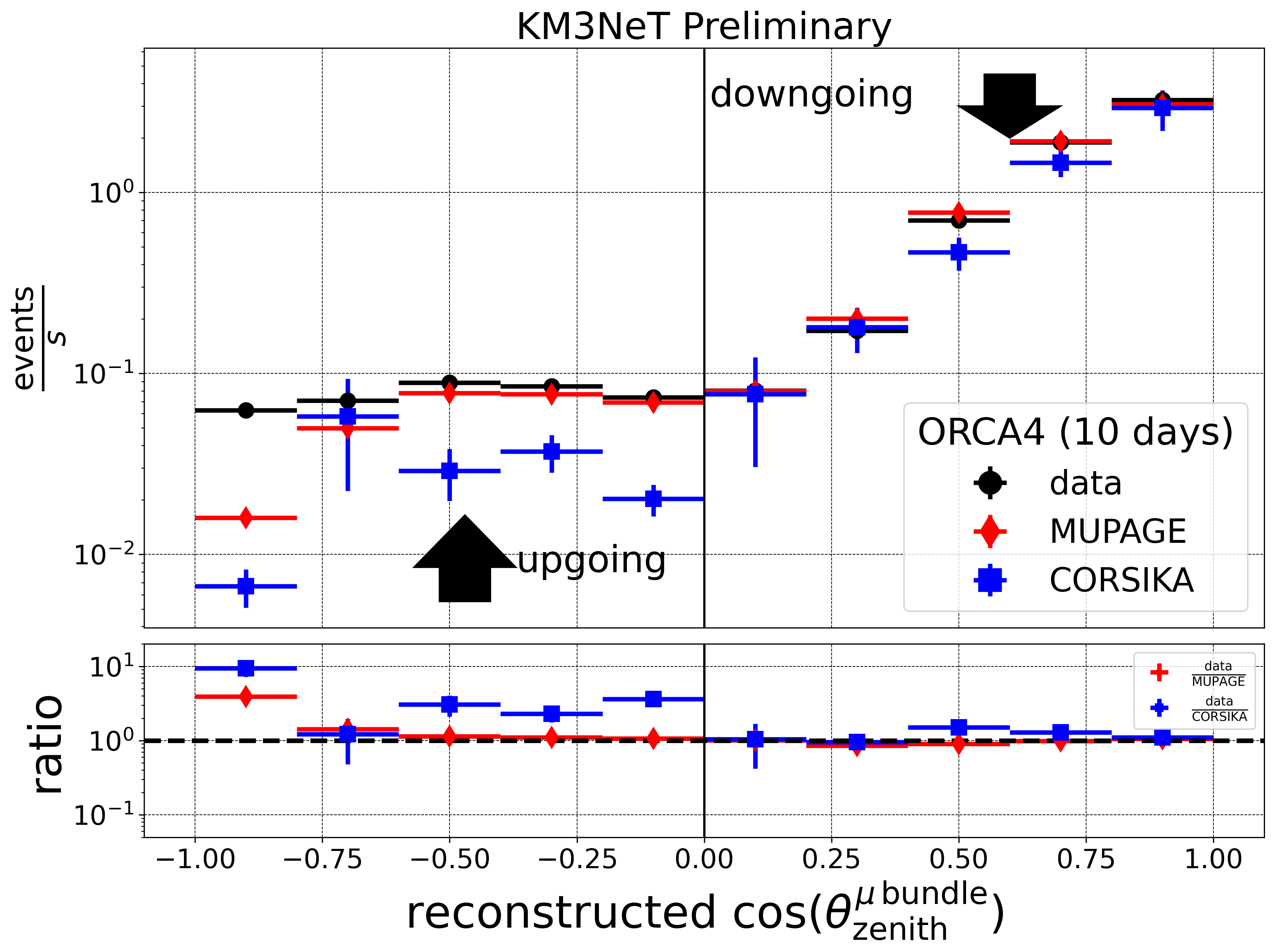}
         \caption{Distribution for ORCA4.}
     \label{fig:zenith-2}
     \end{subfigure}
        \caption{Rate of atmospheric muons as a function of the reconstructed zenith angle for data and MC simulation for the ARCA2 and ORCA4 detector. Most of the upgoing events (all MC events) are downgoing, but are badly reconstructed as upgoing. No quality cuts have been applied to remove the badly reconstructed tracks.}
        \label{fig:zeniths}
\end{figure}

\begin{figure}
     \centering
     \begin{subfigure}[b]{0.49\textwidth}
         \centering
         \includegraphics[width=\textwidth]{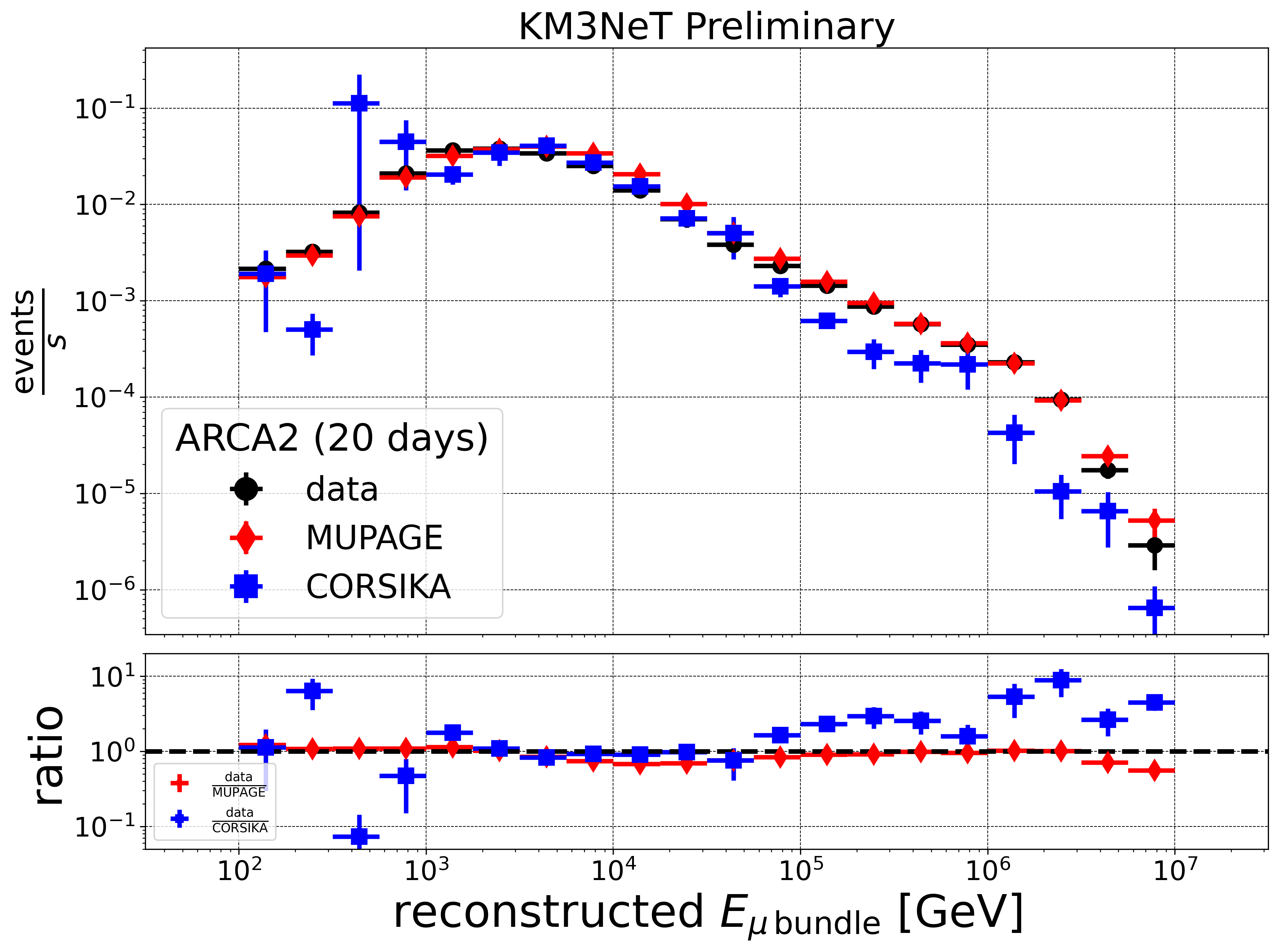}
         \caption{Distribution for ARCA2.}
         \label{fig:energy-1}
     \end{subfigure}
     \hfill
     \begin{subfigure}[b]{0.49\textwidth}
         \centering
         \includegraphics[width=\textwidth]{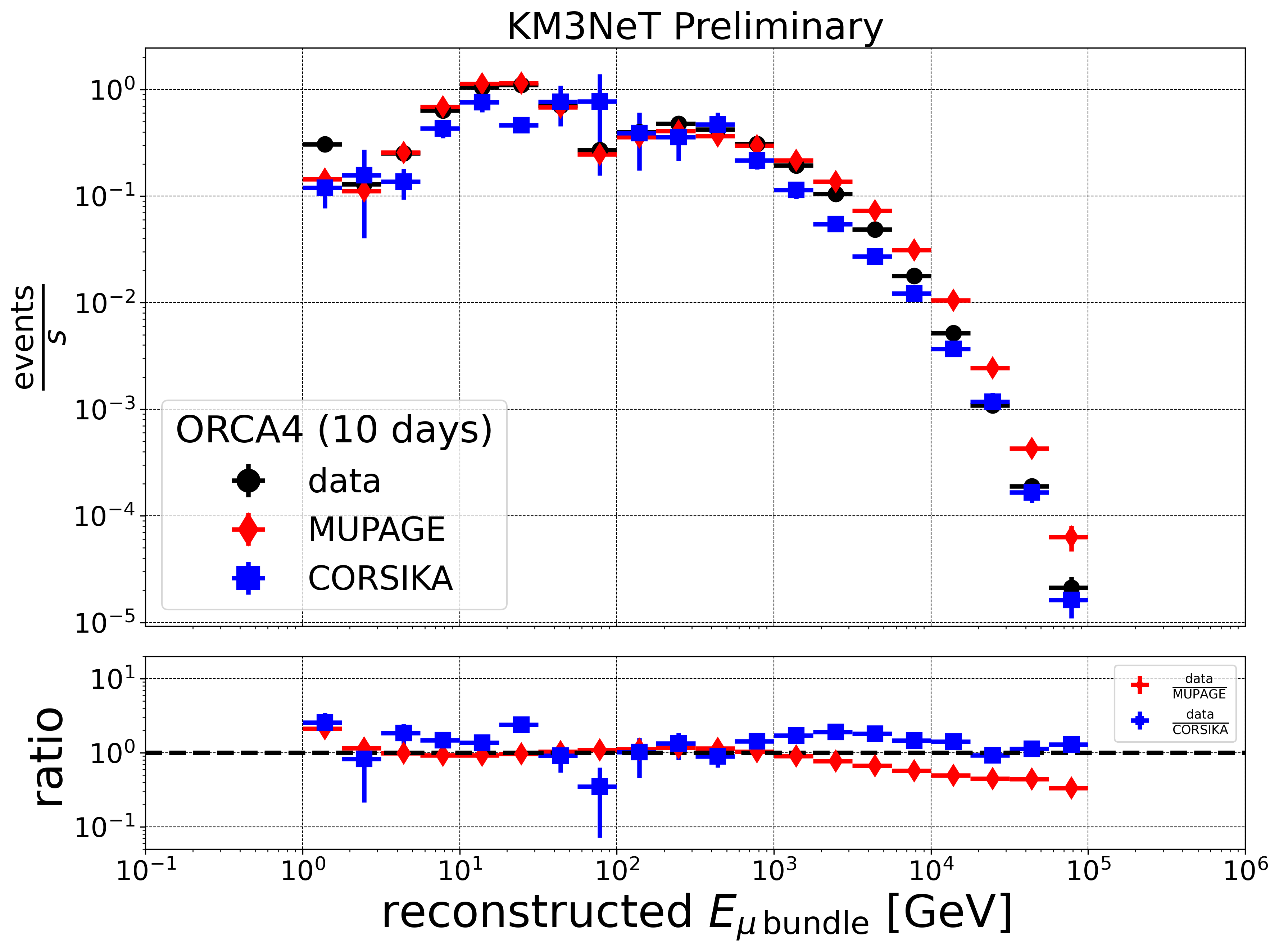}
         \caption{Distribution for ORCA4.}
     \label{fig:energy-2}
     \end{subfigure}
        \caption{Rate of atmospheric muons as a function of the reconstructed energy for data and MC simulation  for the ARCA2 and ORCA4 detector. More details on the energy reconstruction can be found in \cite{reco}.}
        \label{fig:energies}
\end{figure}

A sample of data collected with 2 DUs of the ARCA detector (ARCA2) from the 23th of December 2016 to the 2nd of March 2017 and with 4 DUs of the ORCA detector (ORCA4) from the 23th of July to 16th of December 2019 has been compared with the expectation from the MC based on the same triggering and reconstruction criteria (described in Section \ref{mu_sim}). 

A sample of muon bundles with energies larger than 10 GeV and multiplicity up to 100 tracks was simulated with MUPAGE. The equivalent livetime is close to  the considered detector livetime, about 20\,days for ARCA2 and 10\,days for ORCA4. 
The simulated MUPAGE events processed, as described in Section \ref{mu_sim}, are shown in Figures \ref{fig:zeniths} and \ref{fig:energies} with the red marks. Only statistical errors are shown in the plots.  

A total number of \(2.5\cdot10^9\) showers have been simulated with CORSIKA, using SIBYLL-2.3c to describe the high-energy hadronic interactions \cite{SIBYLL}. Five different species of nuclei were considered as primaries: \(p\), \(He\), \(C\), \(O\) and \(Fe\) with energies between 1\,TeV and 1\,EeV.
The procedure presented in Section \ref{mu_sim} has been followed and events were reconstructed with the same algorithm as for MUPAGE and experimental data. At the end, the events have been weighted according to the GST3 CR composition model \cite{GST}.
The results of the CORSIKA simulation are shown as blue markers in Figures \ref{fig:zeniths} and \ref{fig:energies} and compared with data indicated as black points. The errors plotted are statistical only (\(\Delta x=\sqrt{{\sum}w_{i}^{2}}\), where \(w_{i}\) is the weight of the \(i\)-th MC event). In all plots no systematic uncertainties are considered.

The early results presented here show that the used MC samples reproduce the data with resonable level of agreement. The reconstructed upgoing muons in Figure \ref{fig:zeniths} are misreconstructed downgoing muons. No quality cuts or selection have been applied to remove poorly reconstructed tracks. Note, that the systematic uncertainties, which have not been considered here, are expected to be larger than the statistical errors.
The fact that the MC reproduces the shape of the data, confirms that the KM3NeT simulation chain is reliable and under control.

\section{Prompt muon analysis}\label{prompt-ana}

The atmospheric muon flux is commonly divided into two categories: conventional muons, created mostly in $\pi^{\pm}$ and $K^{\pm}$ decays and prompt muons originating predominantly from decays of short-lived heavy hadrons. With this analysis, the potential of KM3NeT to observe the prompt muon flux is investigated.

The prompt sensitivity study is performed on a CORSIKA MC sample different than the one used for results of Section \ref{data-MC}. The total number of simulated showers is \(5.3\cdot10^6\) and SIBYLL-2.3d is used for high-energy hadronic interactions \cite{SIBYLL-2.3d}. Primary CR energies are simulated in the range from 0.9\,PeV up to 40\,EeV. Lower primary energies are not considered, since already PeV primaries will produce at most 10\,TeV muons, which are well below the region, where the prompt muons are expected to start dominating. At these energies, they will be strongly overwhelmed by the conventional muon flux. 

\begin{figure}
     \centering
     \includegraphics[width=0.8\textwidth]{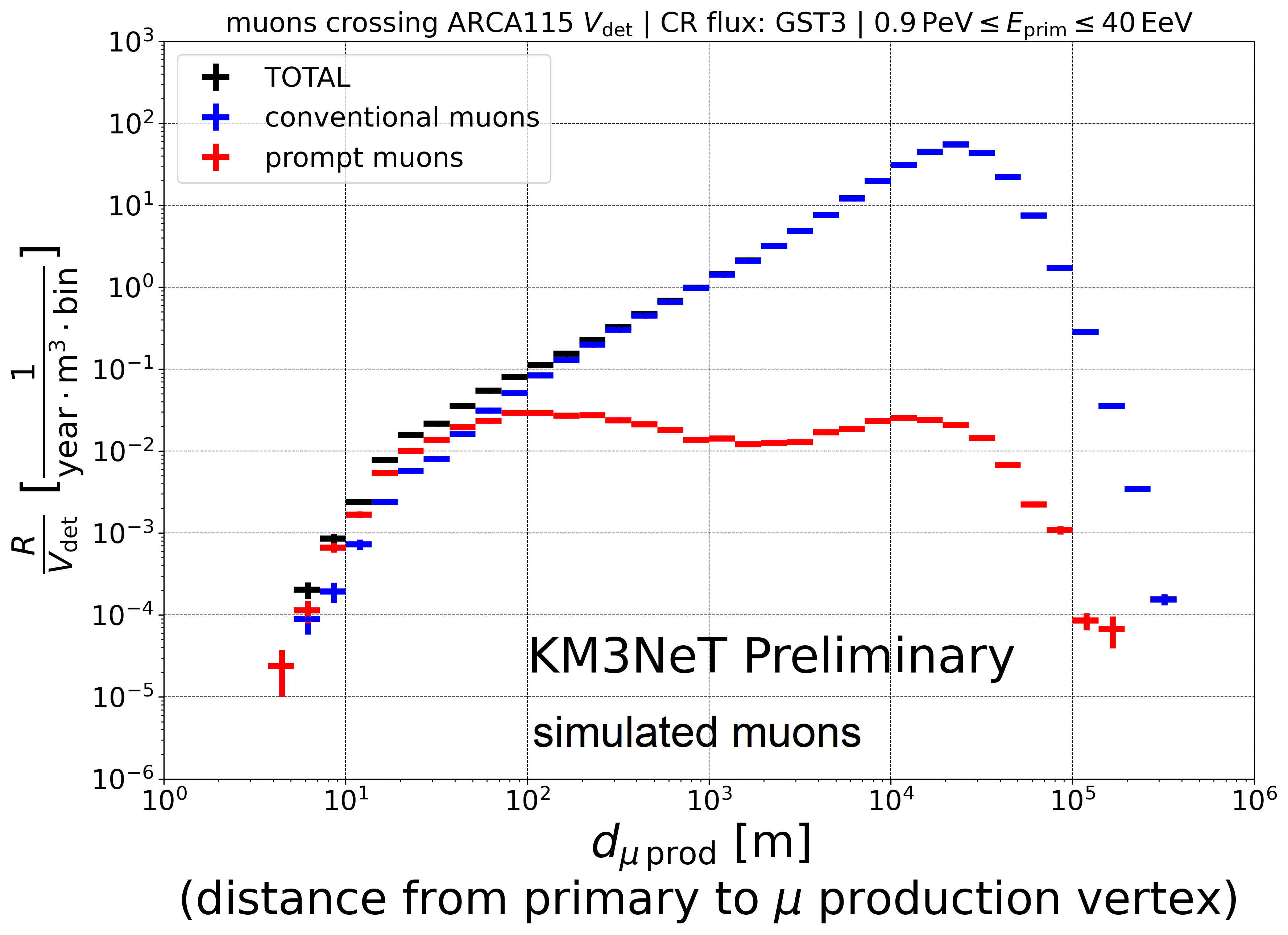}
     \caption{Rate of atmospheric muons per detector volume as a function of the distance between the first interaction of the primary and the point, where the muon was produced. The plot is done for ARCA115 detector based on the CORSIKA Monte Carlo at the detector level without light simulation, triggering or reconstruction.}
     \label{fig:prompt-1}
\end{figure}

\begin{figure}
     \centering
     \includegraphics[width=0.8\textwidth]{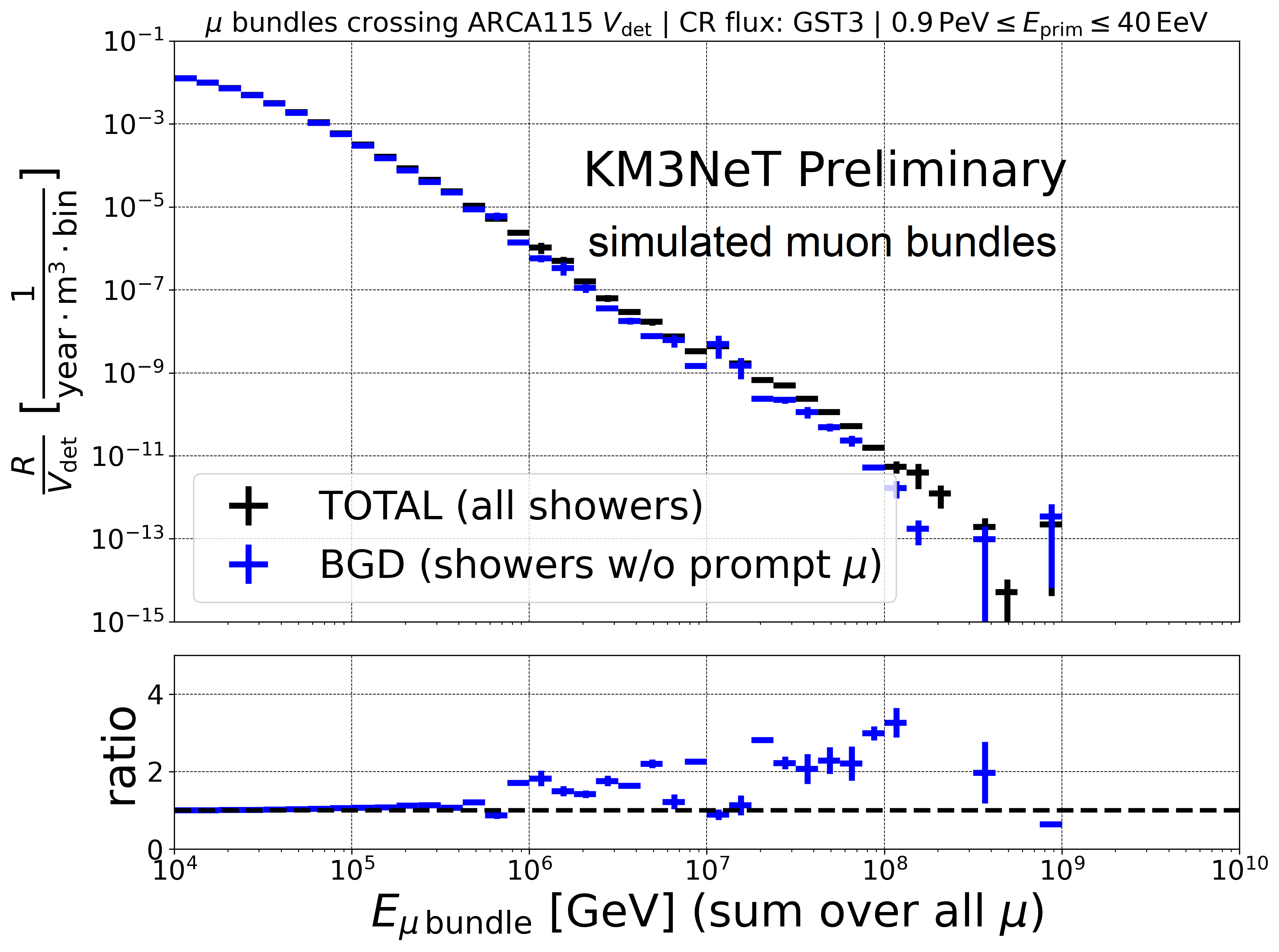}
     \caption{Rate of atmospheric muon bundles per detector volume as a function of the bundle energy (sum of energies of individual muons) for BGD and TOTAL for the ARCA115 detector.}
     \label{fig:prompt-2}
\end{figure}

The prompt muon compontent was defined for the CORSIKA MC events based on two conditions: that all parent particles have lifetimes shorter than the one of $K_{\mathsf{S}}^{0}$ ($8.95\cdot10^{-11}$s) and that there are at most three generations of particles between the primary and the muon. This definition is tested in Figure \ref{fig:prompt-1}, where it is shown how far from the interaction of the primary particle that started the shower, the muon is produced. Very close to the first interaction, there is indeed an excess coming from the prompt muons as expected, which confirms that the definition in use is sensible. The signal (SIG) for the analysis is defined to be muon bundles with at least one prompt muon (according to the introduced definition). Background (BGD) are the bundles with zero prompt muons and TOTAL denotes all muon bundles put together. The distributions of muon bundle energies for BGD and TOTAL are shown in figure \ref{fig:prompt-2} and there is a clear excess above BGD starting around 1\,PeV, although difference is not large. This expected signal is promising and has to be verified at the reconstruction level.

\section{Conclusions}

With the first KM3NeT data, it is possible to compare the measured and the expected muon rate at ARCA and ORCA sites. The agreement between the data and MC is satisfactory and the simulation procedure provides reliable results. 
Nonetheless, there are ongoing efforts to further improve the description of the data.
In the nearest future, the analysis will be repeated for data collected with the present configuration of the detectors: ARCA6 and ORCA6 (ARCA and ORCA each with six DUs installed) and using new MUPAGE and CORSIKA MC productions. A number of improvements is foreseen for the MC chain: optimisation of the muon propagation step, better definition of the geometry of muon propagation, a more accurate calculation of the event weights and addition of the contribution from delta rays to the energy reconstruction, to name a few. A systematic uncertainty study and an increase of the statistical significance of the sample, especially at the highest energies are planned for the CORSIKA MC.

The first results from the prompt muon sensitivity study are enouraging and suggest KM3NeT is sensitive to the prompt component. Figure \ref{fig:prompt-2} presents the expected result for ARCA115  (one building block of the ARCA detector, with 115 DUs). A similar behaviour has also been seen for ORCA115 simulation. A verification at the reconstruction level is necessary to ensure that detectors are efficient enough to extract the prompt signal.
After the complete workflow of the analysis is established, comparison to the results of IceCube \cite{ICECUBE_PROMPT} is planned with the newest prompt muon flux models. Additionally, the muon multiplicity (observed number of muons in a bundle) has been found to be useful observable for the analysis and will be included to further boost the sensitivity.

\section*{Acknowledgements}
This work was supported by the National Centre for Science: Poland grant 2015/18/E/ST2/00758.

\clearpage
\section*{Full Author List: KM3NeT Collaboration}

\scriptsize
\noindent
M.~Ageron$^{1}$,
S.~Aiello$^{2}$,
A.~Albert$^{3,55}$,
M.~Alshamsi$^{4}$,
S. Alves Garre$^{5}$,
Z.~Aly$^{1}$,
A. Ambrosone$^{6,7}$,
F.~Ameli$^{8}$,
M.~Andre$^{9}$,
G.~Androulakis$^{10}$
% ...
\\

\noindent
$^{1}$Aix~Marseille~Univ,~CNRS/IN2P3,~CPPM,~Marseille,~France. \\
$^{2}$INFN, Sezione di Catania, Via Santa Sofia 64, Catania, 95123 Italy. \\
$^{3}$Universit{\'e}~de~Strasbourg,~CNRS,~IPHC~UMR~7178,~F-67000~Strasbourg,~France. \\
$^{4}$Universit{\'e} de Paris, CNRS, Astroparticule et Cosmologie, F-75013 Paris, France. \\
$^{5}$IFIC - Instituto de F{\'\i}sica Corpuscular (CSIC - Universitat de Val{\`e}ncia), c/Catedr{\'a}tico Jos{\'e} Beltr{\'a}n, 2, 46980 Paterna, Valencia, Spain. \\
$^{6}$INFN, Sezione di Napoli, Complesso Universitario di Monte S. Angelo, Via Cintia ed. G, Napoli, 80126 Italy. \\
$^{7}$Universit{\`a} di Napoli ``Federico II'', Dip. Scienze Fisiche ``E. Pancini'', Complesso Universitario di Monte S. Angelo, Via Cintia ed. G, Napoli, 80126 Italy. \\
$^{8}$INFN, Sezione di Roma, Piazzale Aldo Moro 2, Roma, 00185 Italy. \\
$^{9}$Universitat Polit{\`e}cnica de Catalunya, Laboratori d'Aplicacions Bioac{\'u}stiques, Centre Tecnol{\`o}gic de Vilanova i la Geltr{\'u}, Avda. Rambla Exposici{\'o}, s/n, Vilanova i la Geltr{\'u}, 08800 Spain. \\
$^{10}$NCSR Demokritos, Institute of Nuclear and Particle Physics, Ag. Paraskevi Attikis, Athens, 15310 Greece. \\
% ...
$^{55}$Universit{\'e} de Haute Alsace, 68100 Mulhouse Cedex, France. \\
% ...

\end{document}